\documentclass[aps,pra,reprint,nofootinbib,superscriptaddress,twocolumn,showpacs,showkeys,longbibliography,amsmath,amssymb]{revtex4-2}
\usepackage{graphicx}
\usepackage{dcolumn}
\usepackage{bm}
\usepackage{braket}
\usepackage{subfigure}
\usepackage[colorlinks,bookmarks=false,citecolor=blue,linkcolor=red,urlcolor=blue]{hyperref}
\usepackage[english]{babel}
\usepackage{changes}

\begin{document}
\preprint{APS/123-QED}
	\title{Fractional Thouless pumping of solitons: a unique manifestation of bulk-edge correspondence of nonlinear eigenvalue problems}
	\author{Chenxi Bai}
	\affiliation{Department of Physics, Zhejiang Normal University, Jinhua 321004, China}
	\author{Zhaoxin Liang}\email[Corresponding author:~] {zhxliang@zjnu.edu.cn}
	\affiliation{Department of Physics, Zhejiang Normal University, Jinhua 321004, China}

\begin{abstract}
Recent foundational studies have established the bulk-edge correspondence for nonlinear eigenvalue problems using auxiliary eigenvalues $\hat{H}\Psi=\omega S(\omega)\Psi$, spanning both linear~[\href{https://doi.org/10.1103/PhysRevLett.132.126601}{ T. Isobe {\it et al.,} Phys. Rev. Lett. 132, 126601 (2024)}] and  nonlinear~[\href{https://doi.org/10.1103/PhysRevA.111.042201}{ Chenxi Bai and Zhaoxin Liang, Phys. Rev. A. 111, 042201 (2025)}] Hamiltionians. This progress prompts a fundamental question: Can eigenvalue nonlinearity generate observable physical phenomena absent in conventional approaches ($\hat{H}\Psi=E\Psi$)? In this work, we address this question by demonstrating the first uniquely nonlinear manifestation of the bulk-edge correspondence: fractional Thouless pumping of solitons. Through systematic investigation of nonlinear Thouless pumping in an extended Rice-Mele model incorporating next-nearest-neighbor (NNN) couplings, we uncover that NNN interaction parameters can induce fractional topological phases—even in the presence of quantized topological invariants as predicted by conventional linear approaches. Crucially, these fractional phases are naturally explained within the auxiliary eigenvalue framework, directly linking nonlinear spectral characteristics to the bulk-boundary correspondence. Our findings reveal novel emergent phenomena arising from the interplay between nonlinearity and NNN couplings, providing key insights for the design of topological insulators and the controlled manipulation of quantum edge states in nonlinear regimes.
\end{abstract}
	
\maketitle
\section{Introduction}
The bulk-edge correspondence stands as a cornerstone of topological matter~\cite{Eschrig2011, Ozawa2019, Huber2016, Haldane2008, Lu2014}, establishing an intimate link between a system's bulk topology and the emergence of robust boundary states~\cite{Amin2018, Storkanova2021, Mezil2017, Hatsugai1993, 2Hatsugai1993} in diverse systems such as photonic systems~\cite{Raghu2008, Wang2008, Wu2015, Takahashi2018,  Yasutomo2020, Yuto2022}, classical wave systems~\cite{Roman2016, Takahashi2019, Liu2020, Makino2022, Hu2022, Knebel2020, Yoshida2021}, acoustic metamaterials~\cite{Yang2015}, and even biological networks~\cite{Kawaguchi2017}. Central to this framework are topological invariants~\cite{Xiao2010, Ghatak2019, Song2019, Yao2018, Kariyado2015, Citro2023}—Chern numbers, $\mathbb{Z}_2$ indices, and their generalizations—which serve as fundamental markers of bulk topology. These invariants are computed via integration over the Brillouin zone using methods like Wilson loops~\cite{Alexandradinata2014} or K-theory~\cite{Teo2010} or real-space wavefunction analysis, derived from the Schrödinger equation $\hat{H}\Psi = E\Psi$ and its linear eigenvalue solutions. In non-Hermitian systems~\cite{Longhi2019, Gong2018, Ghatak2020, Xiao2020, Zeuner2015, Li2023, Haldane1981}, this paradigm undergoes a profound shift: the standard Brillouin zone must be replaced by a generalized Brillouin zone (GBZ)~\cite{Guo2021, Ji2024, Yang2020}, a complex momentum contour, to reconcile bulk topology with skin effects~\cite{Li2020, McDonald2018, Kunst2018} and predict edge states. The universality of bulk-edge correspondence finds its deepest expression in topological index theorems (e.g., Atiyah--Singer~\cite{Atiyah1968, 2Atiyah1968, 3Atiyah1968, 4Atiyah1968, 5Atiyah1968, Witten2016, Freed2021}), manifested through physical boundary phenomena. Future work can extend this foundation to unify bulk-edge physics across interacting~\cite{Jurgensen2021,Jurgensen2023,Viebahn2024,Cao2024,Cao2024a,Cao2025}  disordered, and driven non-equilibrium systems~\cite{Rudner2020, Kawabata2019, Flaschner2018}—a grand challenge demanding new mathematical tools.

Conceptually going beyond above conventional approaches, a recent breakthrough~\cite{Isobe2024} has forged a connection between bulk-edge correspondence and eigenvalue nonlinearity in linear Hamiltonians via the auxiliary eigenvalue formulation  
$\hat{H}\Psi = \omega S(\omega)\Psi$. This framework elucidates how topological edge states of auxiliary eigenstates manifest as physical edge states under weak yet finite nonlinearity (requiring monotonic $\omega$-dependence), rigorously extending the bulk-edge correspondence to nonlinear eigenvalue problems~\cite{Goblot2019, Ezawa2022, Ezawa2022_2, Sone2022, Pernet2022, Jezequel2022, Sato2023, Kazuki2024, Schindler2025}. Building upon this foundation, our work~\cite{Bai2025} further generalizes the auxiliary eigenvalue approach to intrinsic nonlinear Hamiltonians, revealing an anomalous bulk-edge correspondence inherently tied to these nonlinear eigenvalues. Collectively, Refs.~\cite{Isobe2024,Bai2025} establish a unified description of bulk-edge physics across linear and nonlinear systems through the lens of nonlinear eigenvalue problems. This raises a pivotal question: Can this framework predict observable phenomena inaccessible to conventional linear bulk-edge paradigms?

\begin{figure}
\includegraphics[width=1\columnwidth]{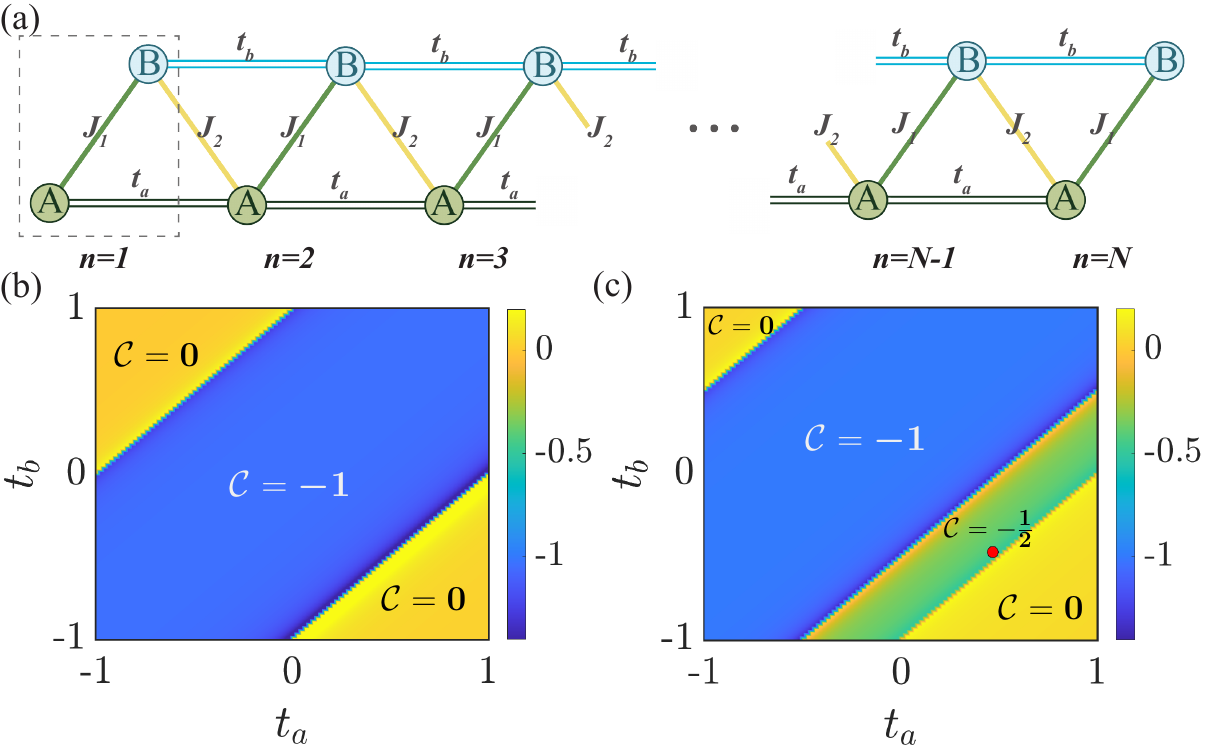}
\caption{\label{fig:1} (a) Schematic diagram of the nonlinear extended RM model. The dotted box indicates the unit cell composed of sublattices A and B.
(b) Chern number $\mathcal{C}$ topological phase diagram of the linear extended RM Hamiltonian (\ref{H}) with $g=0$ under $H_{\text{ERM}}\Psi=E\Psi$.
(c) Chern number $\mathcal{C}$ topological phase diagram of the linear extended RM Hamiltonian (\ref{H}) with $g=0$ under $H_{\text{ERM}}\Psi=\omega S(\omega)\Psi$. The red dot represent the fractional Chern number of $\mathcal{C}=-\frac{1}{2}$ in specific NNN coupling parameter regimes of $t_a$ and $t_b$. (b,c) The parameters are given as $J=1$, $\delta=0.5$, $\Delta=1$, $T=2000\pi$, and $\omega_{\text{d}}=10^{-3}$}.
\end{figure}

This work explores bulk-edge correspondence under eigenvalue nonlinearity $\hat{H}_{\text{ERM}}\Psi=\omega S(\omega)\Psi$, yielding physical observables inaccessible to conventional linear correspondence ($\hat{H}_{\text{ERM}}\Psi=E\Psi$). We implement this through nonlinear topological phase diagrams in an extended Rice-Mele model with next-nearest-neighbor (NNN) couplings $\{t_a, t_b\}$ and nonlinearity strength $g$. First, we present topological phase diagrams for both linear ($H_{\text{ERM}}\Psi = E \Psi$) and nonlinear ($\hat{H}_{\text{ERM}}\Psi=\omega S(\omega)\Psi$) regimes, showing Chern number $\mathcal{C}$ evolution (Figs.~\ref{fig:1}(b),(c)). Crucially, we identify fractional Chern numbers in specific parameter regimes (Fig.~\ref{fig:1}(c)).  We then investigate soliton pumping dynamics by tracking center-of-mass evolution through adiabatic cycles by the numerical methods of
dynamically evolved soltions and instantaneous solitons with the theoretical framework of nonlinear eigenvalue problem (see Figs. \ref{fig:2}-\ref{fig:4} ) and anomalous eigenvalue's nonlinearity problem (see Figs. \ref{fig:5}-\ref{fig:7})  respectively. 
Both approaches demonstrate how dynamically tuned NNN hoppings and nonlinearity $g$ drive integer-to-fractional Chern phase transitions (Figs.~\ref{fig:4}(e),~\ref{fig:7}(e)). Finally, by constructing a 2D momentum-cyclic parameter space, we define topological invariants correlated with nonlinear eigenvalue bifurcations, obtaining a Haldane-model-like phase diagram. Our results predict directly observable fractional nonlinear Thouless pumping in photonic waveguides or cold-atom systems.

This paper is structured as follows: In Section~\ref{Section2}, we present a comprehensive introduction to the Hamiltonian formulation and Schrödinger equation governing the nonlinear extended RM model. Section~\ref{Section3} analyzes the energy spectra and phase diagrams of this model under both linear and nonlinear eigenvalue conditions, with a focus on how nonlinearity strength and NNN coupling parameters influence these properties. In Section~\ref{Section4}, we explore the bulk-edge correspondence within the nonlinear eigenvalue regime of anomalous states, elucidating the role of key model parameters in mediating topological phenomena. Finally, Section~\ref{Section5} summarizes our principal findings and their implications for future research.

\section{Nonlinear extended Rice-Mele model}
\label{Section2}
The model system studied in this work consists of a one-dimensional interacting bosonic chain with $N$ dimer units. As illustrated in Fig. \ref{fig:1} (a),  each unit cell contains two lattice sites (denoted as A and B respectively). Within the mean-field approximation, the dynamics of our bosonic dimer chain is effectively captured by a nonlinear extension of the Rice-Mele Hamiltonian, incorporating both  NNN couplings,~\cite{Rice1982, Lin2020, Tuloup2023, Cerjan2020, Kofuji2024, Hayward2018, Walter2023, Viebahn2024, Hayward2021, Allen2022,  Kofuji2024, Abhishek2023} 
\begin{equation}
H_{\text{ERM}} = H_{\text{RM}} + H_{\text{NNN}},\label{H}
\end{equation}
with
\begin{eqnarray}  
H_{\text{RM}}&= & \sum_{n=1}^{N}\left(J_1\Psi _{n,A}^{*}\Psi _{n,B}+J_2\Psi _{n,B}^{*}\Psi _{n+1,A}+\text{H.c.}\right)\nonumber\\
 & -&\Delta\cos\left(\omega_{\text{d}} t\right)\sum_{n=1}^{N}\left(|\Psi _{n,A}|^2-|\Psi _{n,B}|^2\right)\nonumber\\
& - & \frac{g}{2}\sum_{n=1}^{N}\left(\left|\Psi_{n,A}\right|^{4}+\left|\Psi_{n,B}\right|^{4}\right),\label{NRM}\\
\!\!\!\!H_{\text{NNN}}\!\!\! &=& \!\!\sum_{n=1}^{N-1} \left( t_a \Psi _{n,A}^{*}\Psi _{n+1,A}+t_b \Psi _{n,B}^{*}\Psi _{n+1,B} \!+\! \text{H.c.} \right). \label{NNN} 
\end{eqnarray}  

In above Eqs. (\ref{NRM}) and (\ref{NNN}), each unit cell consists of two sublattice sites labelled by A and B, and the collective wavefunction $\Psi_n = [\psi_{n,A}, \psi_{n,B}]$ encapsulates the amplitude on the $n$-th dimer. The intracell and intercell couplings are $J_1 = -J - \delta \sin(\omega_{\text{d}} t)$ and $J_2 = -J + \delta \sin(\omega_{\text{d}} t)$ respectively, where $J$ sets the uniform hopping amplitude. The staggered sublattice potential $\Delta \cos(\omega_{\text{d}} t)$ introduces an energy offset between sites, which is a periodically modulates dimerization pattern, while nonlinear interactions are introduced via a Kerr-type term with strength $g$, favoring self-localization~\cite{Tuloup2023}. The NNN Hamiltonian $H_{\text{NNN}}$ contains hopping amplitudes $t_a$ and $t_b$ along sublattices A and B. Crucially, the model reduces to the standard nonlinear Rice-Mele Hamiltonian~\cite{Tuloup2023} when $t_a = t_b = 0$. Finally, we note that Hamiltonian (\ref{NRM}) provides a direct theoretical framework for recent experiments~\cite{Viebahn2024} investigating interaction-induced Thouless pumping in dynamical superlattices or pulsed light propagation in waveguide arrays~\cite{Tuloup2023}.

\subsection{Topological phase diagram for linear and nonlinear eigenvalue problems}

The topological physics of the nonlinear extended Rice-Mele model [Hamiltonian (\ref{H})] is governed by five parameters: nearest-neighbor couplings $J_1$, $J_2$, NNN hoppings $t_a$, $t_b$, and interaction strength $g$. The emphasis and value of this work is to demonstrate that eigenvalue nonlinearity [Refs.~\cite{Isobe2024,Bai2025}] enables novel bulk-edge correspondence, yielding physical observables inaccessible in linear systems ($H_{\text{ERM}}\Psi=E\Psi$).

We first compute the topological phase diagram for the linear extended Rice-Mele model [Hamiltonian (\ref{H}) with $g=0$]. Under periodic boundary conditions, the momentum-space wavefunctions for each sublattice are obtained by discrete Fourier transformation:
$|\Psi_A(k,t)\rangle = 1/\sqrt{N} \sum_{j=1}^N e^{i k j} |\Psi_{j,A}(t)\rangle$ and  $|\Psi_B(k,t)\rangle = 1/\sqrt{N} \sum_{j=1}^N e^{i k j} |\Psi_{j,B}(t)\rangle$,
with $k = 2\pi m/N$ ($m=0,1,\dots,N-1$). The total wavefunction is expressed as a two-component spinor $|\Psi(k,t)\rangle = (|\Psi_A(k,t)\rangle, |\Psi_B(k,t)\rangle)^T$, and then the Hamiltonian $H_{\text{ERM}}(k,t)$ can be written in the form of $H_{\text{ERM}}(k,t) = \langle\Psi(k,t)|h(k,t)|\Psi(k,t)\rangle$. Here, the $h(k,t)$ can expressed in the form of $h(k,t)=h_{\text{I}}I+d(k,t)\cdot \sigma$ with $\text{I}$ being the identity matrix, $\sigma=\left(\sigma_x,\sigma_y,\sigma_z\right)$ representing the Pauli matices, and $h_{\text{I}}=(t_a+t_b)\cos k$, $d_x=(J_1+J_2\cos k)$, $d_y=J_2\sin k$, and $d_z=\left(t_{a}-t_{b}\right)\cos k-\Delta\cos(\omega_{\text d}t)$. This Hamiltonian of $h(k,t)$ has broken chiral symmetry and inversion symmetry, and exhibit no other protecting symmetry. The robust quantized pumping arise from the nontrivial Chern number associated with band structures in the combined time-momentum space. The Chern number, defined over the 2D torus $(k,\theta)$ with $\theta=\omega_{\text{d}} t\in [0,2\pi]$, does not rely on symmetry and serves as a topological invariant that ensures the quantization of the pumped charge per cycle, as long as the energy gap remains open during the adiabatic evolution. In more details, the Chern number for the first non-degenerate band is defined as the integral of the Berry curvature over the Brillouin zone~\cite{Xiao2010,Isobe2024}: $\mathcal{C} = \frac{1}{2\pi} \int_0^{2\pi} dk \int_0^T dt  \left[ \nabla_{\vec{R}} \times \mathcal{A}(\vec{R}) \right]$, where $\vec{R} = [0, 2\pi] \times [0, T]$ and $\mathcal{A}(\vec{R}) = i \langle \Psi(k,t) | \nabla_k | \Psi(k,t) \rangle$ is the Berry connection.

Next, there are two approaches to defining the eigenvalue problem for the Hamiltonian (\ref{H}) \cite{Isobe2024,Bai2025}, i. e. one is based on the conventional approaches of $H_{\text{ERM}}\Psi = E \Psi$; The other
is referred as to the the bulk-edge correspondence of nonlinear eigenvalue problems of linear Hamiltonian (\ref{H}) with $g=0$ defined as \cite{Isobe2024,Bai2025}
\begin{equation}
H_{\text{ERM}} (k,t)\Psi = \omega S(\omega)\Psi.\label{LinerP}
\end{equation}  
We begin by defining the auxiliary matrix $P\left(\omega,k\right)=H_{\text{ERM}}\left( k,t\right)-\omega S\left(\omega\right)$, whose kernel, $P\left(\omega,\mathbf{k}\right)\varPsi=0$, reproduces the nonlinear eigen-problem stated in Eq. (\ref{LinerP}). To elucidate the bulk-edge correspondence encoded in this equation, we promote the equality to an eigen-problem $P\left(\omega,k\right)\varPsi = \lambda\varPsi$, with $\lambda \in \mathbb{R}$ serving as a bookkeeping parameter. Physical relevance is restricted to zero-eigenvalue subspace $\lambda=0$, so the central task reduces to solving $P\left(\omega,k\right)\varPsi = \lambda\varPsi$ at $\lambda=0$. The overlap matrix $S (\omega)$ entering Eq. (\ref{LinerP}) is constructed as $S (\omega)=I-M_S\sigma_z$, where $\sigma_z$ donates the Pauli z-matrix and $M_{S}\left(\omega\right)=M_{1}\tanh\left(\omega t\right)/\omega$ encodes the nonlinear $\omega$-dependence.


Finally, in Figs. \ref{fig:1}(b) and (c), we present the topological phase diagrams depicting the Chern number $\mathcal{C}$ for the linear eigenvalue problem ($H_{\text{ERM}}\Psi = E \Psi$) and the nonlinear eigenvalue problem (Eq. \ref{LinerP}), respectively, both evaluated at $g=0$ for $H_{\text{ERM}}$. Within the linear framework ($H_{\text{ERM}}\Psi=E\Psi$), the value of the Chern number $\mathcal{C}$ remains quantized at $-1$ (Fig.~\ref{fig:1}(b)). In contrast, the nonlinear eigenvalue problem $H_{\text{ERM}}\Psi = \omega S(\omega)\Psi$ yields fractional $\mathcal{C}$, exemplified by $\mathcal{C} = -\frac{1}{2}$ at $t_a=0.5$, $t_b=-0.5$ (red dot, Fig.~\ref{fig:1}(c)). Our subsequent goal is to demonstrate how the bulk-edge correspondence, modified by the eigenvalue nonlinearity evident in Fig. \ref{fig:1}(c), gives rise to the observable physical phenomenon of fractional soliton pumping. This phenomenon is absent in the linear case shown in Fig. \ref{fig:1}(b).
\\
\subsection{Nonlinear extended Schr\"odinger equation}
Our strategy for demonstrating the physical manifestations of the fractional Chern number within the nonlinear framework (Fig. \ref{fig:1}(c)) is inspired by the concept of nonlinear Thouless pumping \cite{Jurgensen2023,Jurgensen2023,Qidong2022,Cao2024,Cao2025}. Specifically, for the nonlinear extended RM Hamiltonian (\ref{H}) with $g\neq 0$, the nonlinearity induces quantization of transport via soliton formation and non-uniform band occupation. Consequently, solitons are expected to undergo fractional pumping in the scenario depicted in Fig. \ref{fig:1}(c).

To study soliton dynamics, we derive the equations of motion for the nonlinear extended RM Hamiltonian $H_{\text{ERM}} $ by applying the variational principle $i\partial \Psi_j/\partial t=\delta H/\delta \Psi^*_j$, yielding the following discrete nonlinear Schrödinger equations for sites $j= 1,...N-1,N$,
\begin{eqnarray}
&&i\frac{\partial\Psi_{j,A}}{\partial t}	=-\left(J+\delta\sin\omega_{\text{d}} t\right)\Psi_{j,B}-\left(J-\delta\sin\omega_{\text{d}} t\right)\Psi_{j-1,B}\nonumber\\
&&-\left[\Delta\cos\omega_{\text{d}} t+g\left|\Psi_{j,A}\right|^{2}\right]\Psi_{j,A}\!+\!t_a(\Psi_{j-1,A}\!+\!\Psi_{j+1,A}),\label{GP1}\\
&&i\frac{\partial\Psi_{j,B}}{\partial t}	=-\left(J+\delta\sin\omega_{\text{d}} t\right)\Psi_{j,A}-\left(J-\delta\sin\omega_{\text{d}} t\right)\Psi_{j-1,A}\nonumber\\
&&+\left[\Delta\cos\omega_{\text{d}} t-g\left|\Psi_{j,B}\right|^{2}\right]\Psi_{j,B}\!+\!t_b(\Psi_{j-1,B}\!+\!\Psi_{j+1,B}).\label{GP2}
\end{eqnarray}
Inserting the ansatz $\Psi (t)= \Psi e^{i\mu t/\hbar}$ into Eqs. (\ref{GP1}) and (\ref{GP2}) yields a sationary nonlinear eigenvalue problem $\hat{H}_{\text{ERM}}\Psi=\mu\Psi$. Eqs. (\ref{GP1}) and (\ref{GP2}) constitute a set of discrete nonlinear Schrödinger equations, formally equivalent to the Gross-Pitaevskii framework in lattice systems. This formalism effectively models both ultracold-atom systems~\cite{Viebahn2024} and nonlinear photonic lattices under periodic driving~\cite{Tuloup2023}.

\begin{figure*}
\includegraphics[width=1.0\textwidth]{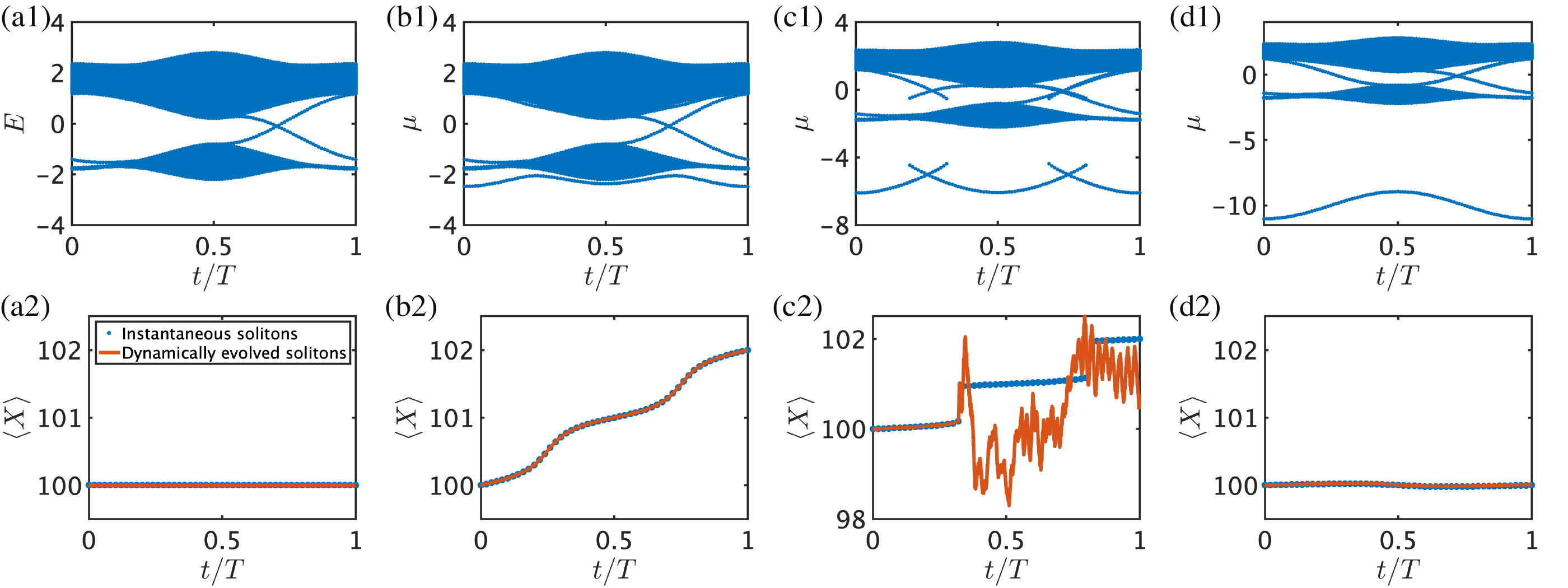}
\caption{\label{fig:2}  (a1)-(d1): Band structures of the nonlinear eigenvalues for the nonlinear extended RM Hamiltonian (\ref{H}), (a2)-(d2): The position expectation value $\langle X\rangle$ of the soliton as a function of time over a single period. The parameters are fixed as $J=1$, $\delta=0.5$, $\Delta=1$, $t_a=0.4$, $t_b=-0.1$, $T=2000\pi$, and $\omega_{\text{d}}=10^{-3}$. (a1)-(d1) [(a2)-(d2)]: The interaction strengths are set to $g=0$, $g=1$, $g=5$ and $g=10$, respectively.}
\end{figure*}

\begin{figure}
\includegraphics[width=1\columnwidth]{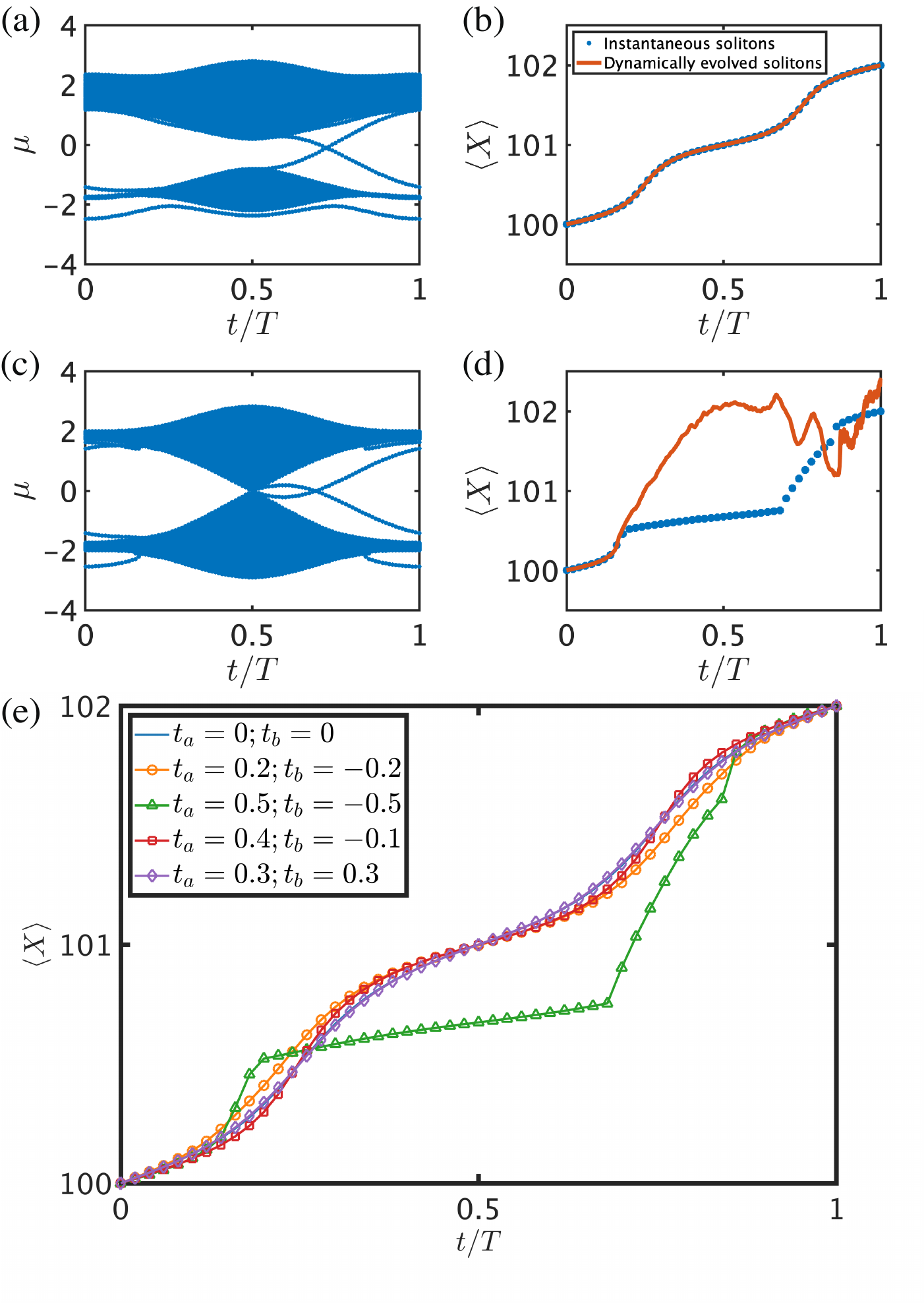}
\caption{\label{fig:3} NNN tunneling control of topological transport: 
(a,c) Nonlinear eigenvalue spectra ($\mu$); 
(b,d,e) Quantized soliton dynamics. 
Fixed parameters: $J=1$, $\delta=0.5$, $\Delta=1$, $g=1$, $T=2000\pi$, $\omega_{\text{d}}=10^{-3}$. 
Tunneling configurations: 
(a,b) $t_a=0.4$, $t_b=-0.1$; 
(c,d) $t_a=0.5$, $t_b=-0.5$.
}
\end{figure}

\section{Fractional nonlinear Thouless pumping of nonlinear extended RM Hamiltonian}
\label{Section3}

In Sec.~\ref{Section2}, we presented the nonlinear extended RM Hamiltonian (\ref{H}) along with the associated discrete nonlinear Schrödinger equations (\ref{GP1}) and (\ref{GP2}).  In  Sec. \ref{Section3}, we plan to investigate the quantized and fractionally-quantized nonlinear Thouless pumping based on Eqs. (\ref{GP1}) and (\ref{GP2}). In more details,  The objective of subsection~\ref{Section3} is to present the band structures of the nonlinear extended RM model as well as the pumping behavior of the soliton, while investigating the effects of the parameters $t_a$ and $t_b$ in the NNN Hamiltonian term (\ref{NNN}). In the subsequent subsection (\ref{S}), we adopt the methodology detailed in Ref.~\cite{Isobe2024} to delve into the anomalous eigenvalue's nonlinearity within the nonlinear extended RM Hamiltonian (\ref{H}) and its corresponding bulk-edge correspondence.

\subsection{Nonlinear eigenvalue of the nonlinear Hamiltonian}\label{nonS}

In this subsection \ref{nonS}, we are interested into both the quantized nonlinear Thouless pumping and fractional nonlinear Thouless pumping predicted by the topological phase diagram in Figs. \ref{fig:1}(b) and (c). This is achieved by numerically solving Eqs.~(\ref{GP1}) and (\ref{GP2}) under a continuous change of the nonlinearity parameter~$ g $ in Eq.~(\ref{H}) from zero to a finite value. All simulations employ a $ 100 $-unit cell lattice ($ 200 $ sites). To ensure numerical reliability, we have systematically verified through size-convergence studies that all simulations in this work demonstrate the system-size independence. The initial soliton state, localized at site~$ n = 0 $, is prepared using the iterative self-consistent algorithm from Refs.~\cite{Bai2025,Tuloup2023} with trial wavefunction $ \Psi_{0} = \text{sech}\left[ \lvert x - 100 \rvert /5 \right] $.  Here, we use two distinct methodologies~\cite{Tuloup2023,Bai2025} to solve  Eqs. (\ref{GP1}) and (\ref{GP2}) numerically, with detailed methodologies presented in Appendix \ref{appendix A}. In brief, the first method is called by dynamically evolved soliton method corresponding to direct time integration of Eqs. (\ref{GP1}) and (\ref{GP2}) using a fourth-order Runge-Kutta algorithm with initial soliton conditions; while, the second method is referred to as instantaneous soliton solution method corresponding to the iterative determination of stationary states through successive approximation, using the previous time-step solution as the initial trial wavefunction.

Now, we are ready to investigate soliton pumping dynamics by monitoring the center-of-mass position through one adiabatic cycle using the two aforementioned methods. This requires calculating the expectation value $\langle X\rangle=\sum_j j|\Psi_j|^2$.

First, we consider how a soliton can be pumped under periodic driving through the lens of nonlinear eigenvalues (see Figs. \ref{fig:2}(b1)-(d1)) corresponding to $H_{\text{ERM}}\Psi=\mu \Psi$ in Table I in Ref. \cite{Bai2025} with the fixed $t_a$ and $t_b$. For baseline comparison, we consider the noninteracting case ($g=0$) of Hamiltonian (\ref{H}), whose band structure is shown in Fig. \ref{fig:2}(a1). Here, solitons are naturally absent due to vanishing nonlinearity, and consequently, no soliton pumping occurs - as expected and demonstrated in Fig. \ref{fig:2}(a2). With the introduction of nonlinearity ($g \neq 0$) in Hamiltonian (\ref{H}), solitons emerge through spontaneous symmetry-breaking bifurcations \cite{Jurgensen2021} and subsequently manifest pumping behavior, as clearly demonstrated in Figs. \ref{fig:2}(b2)-(d2).

The quantized soliton transport in Fig.~\ref{fig:2} originates from the Chern number (see Fig.~\ref{fig:1}(b)) in the noninteracting limit ($g=0$) \cite{Jurgensen2021,Bai2025,Tuloup2023,Qidong2022}. Specifically, under weak nonlinearity (Fig.~\ref{fig:2}(b1)), the soliton occupies the lowest band of Hamiltonian (\ref{H}), whose linear dynamical Chern number equals $-1$ (Fig.~\ref{fig:1}(b)). Consequently, the bulk-edge correspondence for the nonlinear extended RM Hamiltonian (\ref{H}) yields quantized displacement of two lattice sites per cycle (Fig.~\ref{fig:2}(b2)). In contrast, strong nonlinearity (Fig.~\ref{fig:2}(d2)) induces Rabi oscillations between the two lowest bands. The vanishing sum of dynamical Chern numbers ($\mathcal{C}_1 + \mathcal{C}_2 = 0$) then freezes soliton transport. For moderate interactions $g$ (Fig.~\ref{fig:2}(c2)), transport remains quantized when calculated via the instantaneous soliton method. However, dynamically evolved soliton trajectories exhibit position jumps (Fig.~\ref{fig:2}(c2)) due to emergent self-intersecting bands in the intermediate nonlinear regime. This band topology, characterized by loop structures \cite{Wu2001}, is visualized in Fig.~\ref{fig:2}(c1).

We next examine how NNN tunneling amplitudes $t_a$ and $t_b$ in Hamiltonian (\ref{H}) influence quantized soliton transport, maintaining fixed weak nonlinearity ($g=1$). These effects are systematically investigated through Fig.~\ref{fig:3}.
We first analyze the case $t_a = 0.4$, $t_b = -0.1$. As demonstrated in Fig.~\ref{fig:3}(a), gapless edge states persist due to preserved inversion symmetry, resulting in stable soliton oscillations (Fig.~\ref{fig:3}(b)). Crucially, the ground-state soliton exhibits a pumping value of $2$ at this parameter point—consistent with Chern number $ \mathcal{C}=-1$ (corresponding to two atoms per unit cell)—and displays perfect adiabatic following, where dynamically evolved solitons precisely match their instantaneous counterparts. When the absolute values of $t_b$ becomes comparable to $t_a$ ($t_a=0.5$, $t_b=-0.5$), inversion symmetry breaks. This symmetry breaking opens a bandgap in the $\mu$-spectrum (Fig.~\ref{fig:3}(c)) and hybridizes edge states into bulk modes, thereby disrupting the adiabatic pathway as evidenced in Fig.~\ref{fig:3}(d). By selecting distinct NNN tunneling amplitudes $\{t_a, t_b\}$ and superposing their corresponding soliton trajectories as shown in Fig.~\ref{fig:3}(e), we observe that all pumping pathways exhibit perfect adiabatic following \textit{except} along the symmetry-broken line $t_a = -t_b = 0.5$. Crucially, adiabatic breakdown occurs exclusively on this $t_a = -t_b$ trajectory. This reveals how NNN tunneling parameters reconfigure the system's topological phase diagram.

\begin{figure}
\includegraphics[width=1\columnwidth]{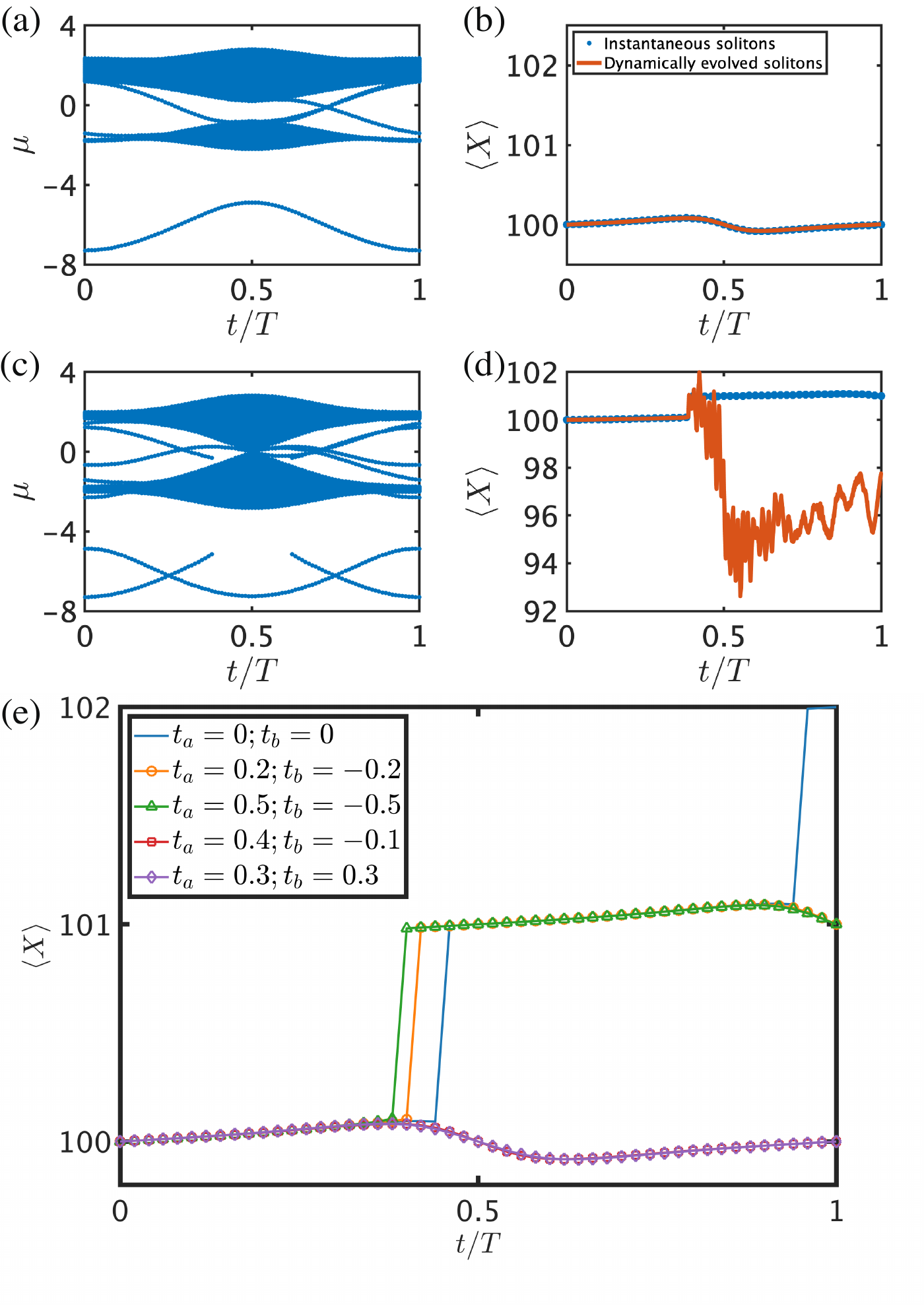}
\caption{\label{fig:4} NNN tunneling modulation of topological phases: 
(a,c) Nonlinear eigenvalue spectra ($\mu$); 
(b,d,e) Quantized soliton transport dynamics. 
Fixed parameters: $J=1$, $\delta=0.5$, $\Delta=1$, $g=6.215$, $T=2000\pi$, $\omega_{\text{d}}=10^{-3}$. 
Parameter configurations:
(a,b) $t_a=0.4$, $t_b=-0.1$;
(c,d) $t_a=0.5$, $t_b=-0.5$.}
\end{figure}

Finally, we demonstrate how the bulk-edge correspondence—modified by nonlinear eigenvalue topology shown in Fig.~\ref{fig:1}(c)—gives rise to fractional soliton pumping. This phenomenon, absent in the linear regime (Fig.~\ref{fig:1}(b)), is investigated at $g=6.215$. As depicted in Fig.~\ref{fig:4}(d), the $t_a = -t_b = 0.5$ configuration maintains adiabatic breakdown (first observed at $g=1$), corresponding to the ground-state soliton's self-intersecting structure in Fig.~\ref{fig:4}(c). Crucially, the transport trajectory obtained via the instantaneous soliton method (blue dotted curve, Fig.~\ref{fig:4}(d)) reveals fractional quantized pumping of the soliton. Conversely, for $t_a = 0.4$, $t_b = -0.1$, self-trapping emerges (Fig.~\ref{fig:4}(a)), suppressing soliton dynamics and consequently halting pumping (Fig.~\ref{fig:4}(b)).

Motivated by the signal of observed fractional Thouless pumping phenomena induced by NNN tunneling $\{t_a, t_b\}$ in Fig.~\ref{fig:4}(d), we are motivated to plot Fig.~\ref{fig:4}(e), which directly demonstrates how the bulk-edge correspondence—modified through nonlinear eigenvalue topology in Fig.~\ref{fig:1}(c)—generates fractional soliton pumping. This constitutes the central finding of our work. The parametric dependence in Fig.~\ref{fig:4}(e) reveals distinct soliton transport regimes: three trajectories exhibit adiabatic breakdown, while configurations $t_a = t_b = 0.3$ and $t_a = 0.4$, $t_b = -0.1$ show dynamical suppression and consequent pumping cessation. Crucially, we observe the crystallization of fractional topological phases in the $t_a = -t_b = 0.2$ and $t_a = -t_b = 0.5$ trajectories. This demonstrates selective reconfiguration of the phase diagram through NNN tunneling, particularly along the $t_a = -t_b$ symmetry line. Systematic comparison of $t_a=-t_b=0.5$ configurations reveals nonlinear control of quantization: at $g=1$ the pumping value remains $2$, while increasing to $g=6.215$ reduces it precisely to $1$.

We remark that the fractional pumping phenomenon reported in Fig.~\ref{fig:4}(e) stems from the introduction of NNN coupling terms and is elucidated within the auxiliary eigenvalue framework in Eq. (\ref{LinerP}), rather than the conventional multi-band Wannier tracking mechanism driven by strong interactions in Ref. \cite{Jurgensen2023}. As explicitly demonstrated in Fig. \ref{fig:3}(e) and Fig. \ref{fig:4}(e), we have carefully verified that merely adjusting the nonlinearity—without including NNN coupling terms ($t_a=t_b=0$)—fails to induce fractional Thouless pumping. This conclusively shows that the predicted fractional pumping cannot be accounted for by the physical picture of solitons coupling to multi-band Wannier functions via strong interactions. Moreover, the authors in Ref. \cite{Tao2025} report the observation of fractional Thouless pumping of solitons in a nonlinear off-diagonal Aubry–Andr{\'e}–Harper model. Notably, this phenomenon in Ref. \cite{Tao2025} arises even though all energy bands of the corresponding linear Hamiltonian are topologically trivial—evidence that nonlinearity can indeed induce fractional Thouless pumping of solitons. In sharp contrast, the linear Hamiltonian associated with the nonlinear eigenvalue problem $H_{\text{ERM}}\Psi = \omega S(\omega)\Psi$ is topologically nontrivial, characterized by a fractional Chern number—for instance, $\mathcal{C} = -\frac{1}{2}$ at $t_a = 0.5$, $t_b = -0.5$ (red dot, Fig.~1(c)).

\begin{figure*}
\includegraphics[width=1\textwidth]{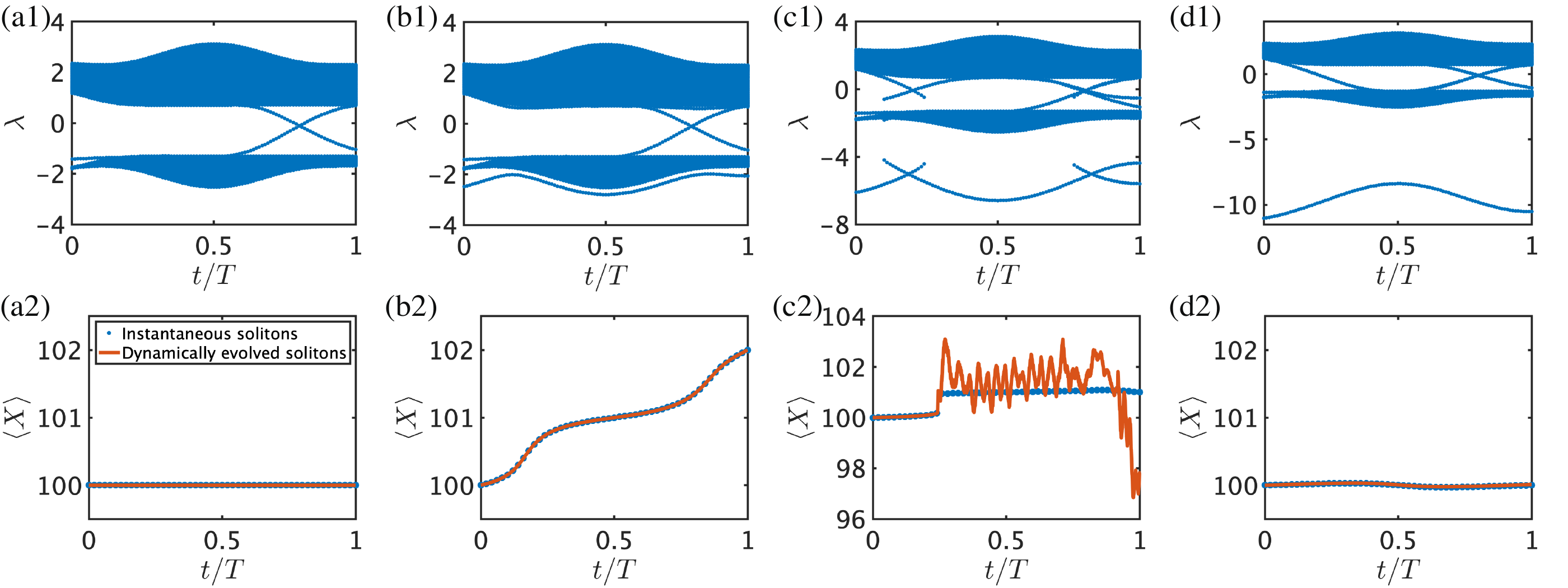}
\caption{\label{fig:5} Eigenvalue's nonlinearity of nonlinear extended RM Hamiltonian and nonlinear Thouless pumping of soliton. (a1)-(d1): The auxiliary $\lambda$-spectrum for the nonlinear extended RM Hamiltonian (\ref{H}). (a2)-(d2): The anticipated position of the nonlinear excitation of soliton as a function of time over one period. The parameters are fixed at  $J=1$, $\delta=0.5$, $\Delta=1$, $t_a=0.4$, $t_b=-0.1$, $T=2000\pi$, and $\omega=\omega_{\text{d}}=10^{-3}$. Specifically, (a1)-(d1) and their corresponding (a2)-(d2) panels represent interaction strengths of $g=0$, $g=1$, $g=5$ and $g=10$, respectively.}
\end{figure*}

\begin{figure}
\includegraphics[width=1\columnwidth]{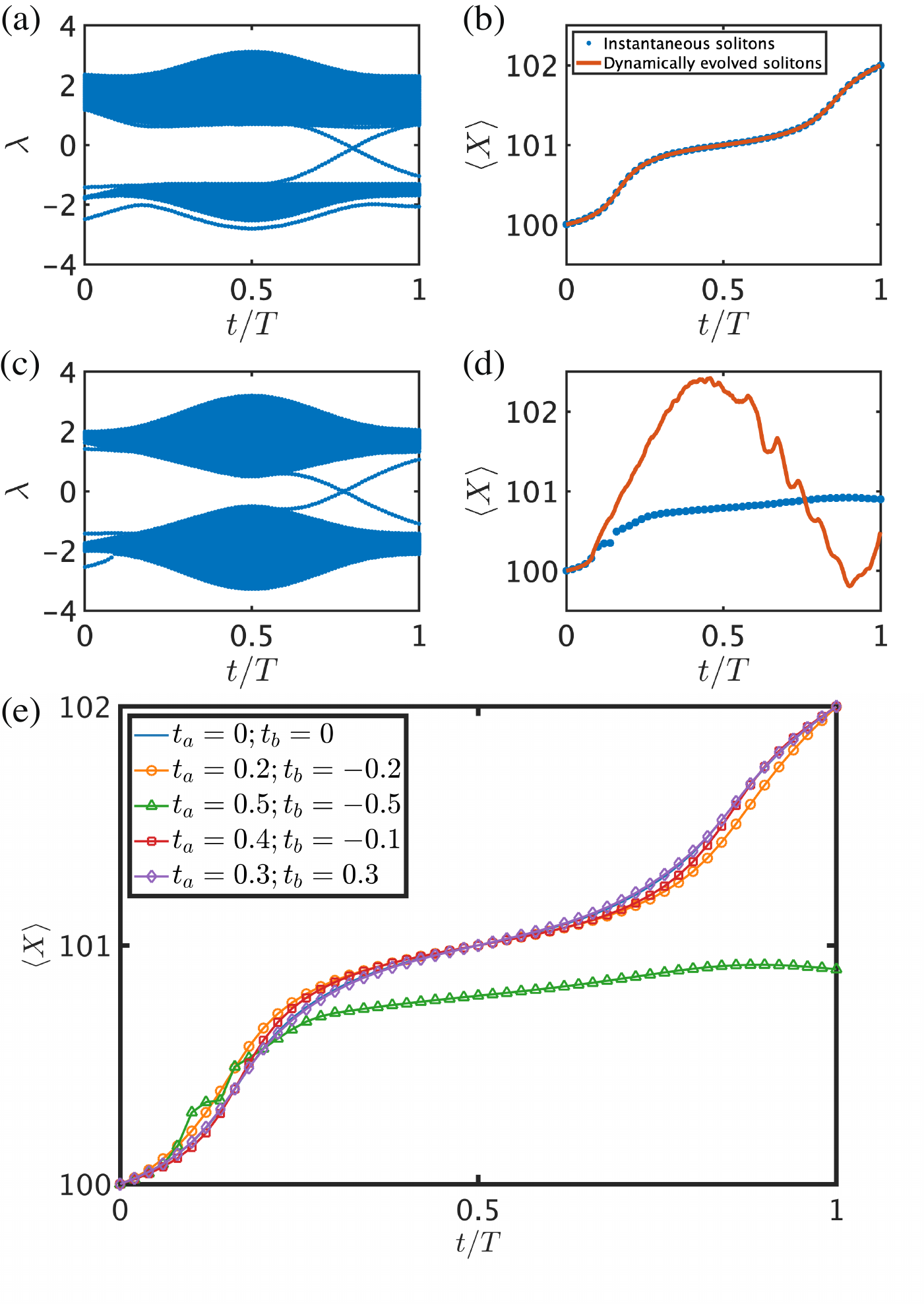}
\caption{\label{fig:6} Effects of parameter $t_a$ and $t_b$ on the auxiliary $\lambda$-spectrum defined in Eq. (\ref{eq:lambda}) for the nonlinear extended RM Hamiltonian (\ref{H}).The parameters are fixed
at $J=1$, $\delta=0.5$, $\Delta=1$, $g=1$, $T=2000\pi$, and $\omega=\omega_{\text{d}}=10^{-3}$. Other parameters are given as (a-b): $t_a=0.4$, $t_b=-0.1$; (c-d): $t_a=0.5$, $t_b=-0.5$. (e): Effects of parameter $t_a$ and $t_b$ on the anticipated position of the nonlinear excitation of soliton as a function of time over one period.}
\end{figure}

\subsection{Anomalous eigenvalue's nonlinearity of nonlinear Hamiltonian}\label{S}

In the preceding subsection (\ref{nonS}), we have analyzed both quantized and fractional nonlinear Thouless pumping for the extended RM Hamiltonian (\ref{H}) within the linear eigenvalue framework. In the subsequent subsection (\ref{S}), we employ the methodology of Refs.~\cite{Isobe2024,Bai2025} to explore anomalous spectral nonlinearities in Hamiltonian (\ref{H}) and their implications for anomalous bulk-edge correspondence first pointed by Ref.~\cite{Bai2025}.

Anomalous eigenvalue nonlinearity generalizes the concept from linear Hamiltonians~\cite{Isobe2024} to nonlinear systems, defined through $\hat{H}_{\text{non}}\Psi = \omega S(\omega) \Psi$. This framework exhibits dual nonlinearity—both Hamiltonian $\hat{H}_{\text{non}}$ and eigenvalue structure $\omega S(\omega)$ become state-dependent—distinguishing it from systems with only wavefunction nonlinearity (e.g., $g|\Psi_n|^2$). Such dual nonlinearity captures quantum systems with anharmonic potentials where energy levels and wavefunctions nonlinearly intertwine, providing a unified description of strongly correlated quantum states.

Our strategy for examining this anomalous eigenvalue nonlinearity in the nonlinear extended RM Hamiltonian (\ref{H}) and bulk-edge correspondence, given the underlying nonlinearity of the eigenvalues, involves the utilization of auxiliary eigenvalues, as outlined below:

(i) The nonlinear extended RM Hamiltonian (\ref{H}) is characterized by an anomalous form of eigenvalue nonlinearity, expressed as \cite{Bai2025}
\begin{equation}
H_{\text{ERM}}\left(\omega_{\text{d}},t\right)\varPsi=\omega S\left(\omega\right)\varPsi.\label{eq:nonlinear}
\end{equation}

Noted that Eq. (\ref{eq:nonlinear}) and the accompanying matrix $S(\omega)$ serve to embed auxiliary eigenvalues into the problem; As a consequence, the topological edge modes of these auxiliary eigenstates are inherited by the physical system, thereby highlighting a subtle interplay between eigenvalue nonlinearity and topology. The overlap matrix $S(\omega)$ in Eq. (\ref{eq:nonlinear}), whose structure varies with the nonlinear parameter $\omega$, is assembled as follows:
\begin{eqnarray}
S\left(\omega\right)=\left(\begin{array}{cccc}
A_{0} & 0 & 0 & 0\\
0 & A_{0} & 0 & 0\\
0 & 0 & \ddots & 0\\
0 & 0 & 0 & A_{0}
\end{array}\right),
\end{eqnarray}
with the diagonal block $A_{0}$ takes the form $A_{0}=1-\sigma_zM_{S}$ and  $\sigma_z$ the $z$-component of Pauli matrix. Here, the scalar function $M_{S}\left(\omega\right)=M_{1}\tanh\left(\omega t\right)/\omega$ governs its dependence on the nonlinear parameter $\omega$.

(ii) We then construct the matrix $P\left(\omega,t\right)$ as follows:
\begin{equation}
P\left(\omega,t\right)=H_{\text{ERM}}\left(\omega_{\text{d}},t\right)-\omega S\left(\omega\right).\label{EquationP}
\end{equation}
Equation (\ref{EquationP}) shows that any solution of $P\left(\omega,\mathbf{k}\right)\varPsi=0$ also satisfies the nonlinear relation in Eq. (\ref{eq:nonlinear}). To explore the bulk-edge correspondence implied by this relation more deeply, we append an auxiliary, real-valued eigenvalue $\lambda$, so that
\begin{equation}
P\left(\omega,t\right)\varPsi = \lambda\varPsi.\label{eq:lambda}
\end{equation}
Although the auxiliary eigenvalue $\lambda$ lacks direct physical meaning in general, the special case $\lambda=0$ is singled out. Thus, the central task becomes solving Eq. (\ref{eq:lambda}) precisely at this point.  Before proceeding, it is worth noting two key distinctions in solving Eqs.~(\ref{LinerP}) and (\ref{eq:nonlinear}). First, Eq.~(\ref{LinerP}) describes the eigenvalue problem for the linear Hamiltonian~{\ref{H}} with $g = 0$, whereas Eq.~(\ref{eq:nonlinear}) corresponds to the nonlinear case with $g \neq 0$. Second, these equations are solved in different representations: Eq.~(\ref{LinerP}) in momentum space and Eq.~(\ref{eq:nonlinear}) in coordinate space.

(iii)  Finally, we turn to investigating both quantized and fractional nonlinear Thouless pumping within the framework of anomalous bulk-edge correspondence \cite{Bai2025} for the extended Rice-Mele Hamiltonian~\eqref{H}. This analysis will proceed by numerically solving the eigenvalue equation~\eqref{eq:lambda}. Furthermore, we will demonstrate topological inheritance from auxiliary to physical systems, showing how topological edge states of the auxiliary eigenstates manifest as physical edge states through numerical solution of Eqs.~(\ref{GP1}) and~\eqref{GP2}. To systematically explore these phenomena, we examine three distinct scenarios.

In the first scenario, we examine the connection between the auxiliary eigenvalue $\lambda$ (Figs.~\ref{fig:5}(a1)-(d1)) obtained from Eq.~(\ref{eq:lambda}) and the soliton transport dynamics (Figs.~\ref{fig:5}(a2)-(d2)) by fixing the $t_a=0.4$ and $t_b=-0.1$.
Specifically, at $g=0$, the pump transport oscillates near the initial point, as shown in Fig.~\ref{fig:5}(a2). As the interaction strength increases to $g=1$ and $g=5$, the system demonstrates pump transport phenomena. When $g=1$, the pump transport value over one period remains 2, as depicted in Fig.~\ref{fig:5}(b2). Furthermore, when the interaction strength continues to rise to $g=5$, the dynamical evolution of the soliton wave within a period is significantly reduced, and the pumping value decreases from 2 to 1, as depicted in Fig.~\ref{fig:5}(c2).
Consistent with the original nonlinear extended RM model, when $g=10$, the pumping ceases to operate in the nonlinear extended RM model with the nonlinear eigenvalue problem, as illustrated in Fig.~\ref{fig:5}(d2).

In the second scenario, we proceed to investigate how NNN tunnelings of  $t_a$ and $t_b$ modify the soliton's transport predicted by Fig. \ref{fig:5}. We first examine the case $t_a = 0.4$, $t_b = -0.1$ and $g=1$. As depicted in Figs.~\ref{fig:6}(a), the gapless edge states persist due to the preservation of inversion symmetry, which corresponds to the stable soliton oscillations shown in Figs.~\ref{fig:6}(b). Similar to the case with linear eigenvalues, the pumping value here is 2, and there is perfect adiabatic following. When $t_a = -t_b = 0.5$, the ground-state soliton exhibits a self-intersecting structure, as illustrated in Figs.~\ref{fig:6}(c). This self-intersecting band structure leads to the breakdown of the adiabatic path, as shown in Figs.~\ref{fig:6}(d). Moreover the results in Figs.~\ref{fig:6}(e)demonstrate that NNN parameters $\{t_a, t_b\}$ actively reconfigure the nonlinear system's phase diagram—a capability absent in linear eigenvalue systems. Specifically, the NNN parameters of $t_a = -t_b=0.5$ (see the green line Figs.~\ref{fig:6}(e)) induces a fractional topological phase transition distinct from linear mechanisms in Fig.~\ref{fig:1}(b).

\begin{figure}
\includegraphics[width=1\columnwidth]{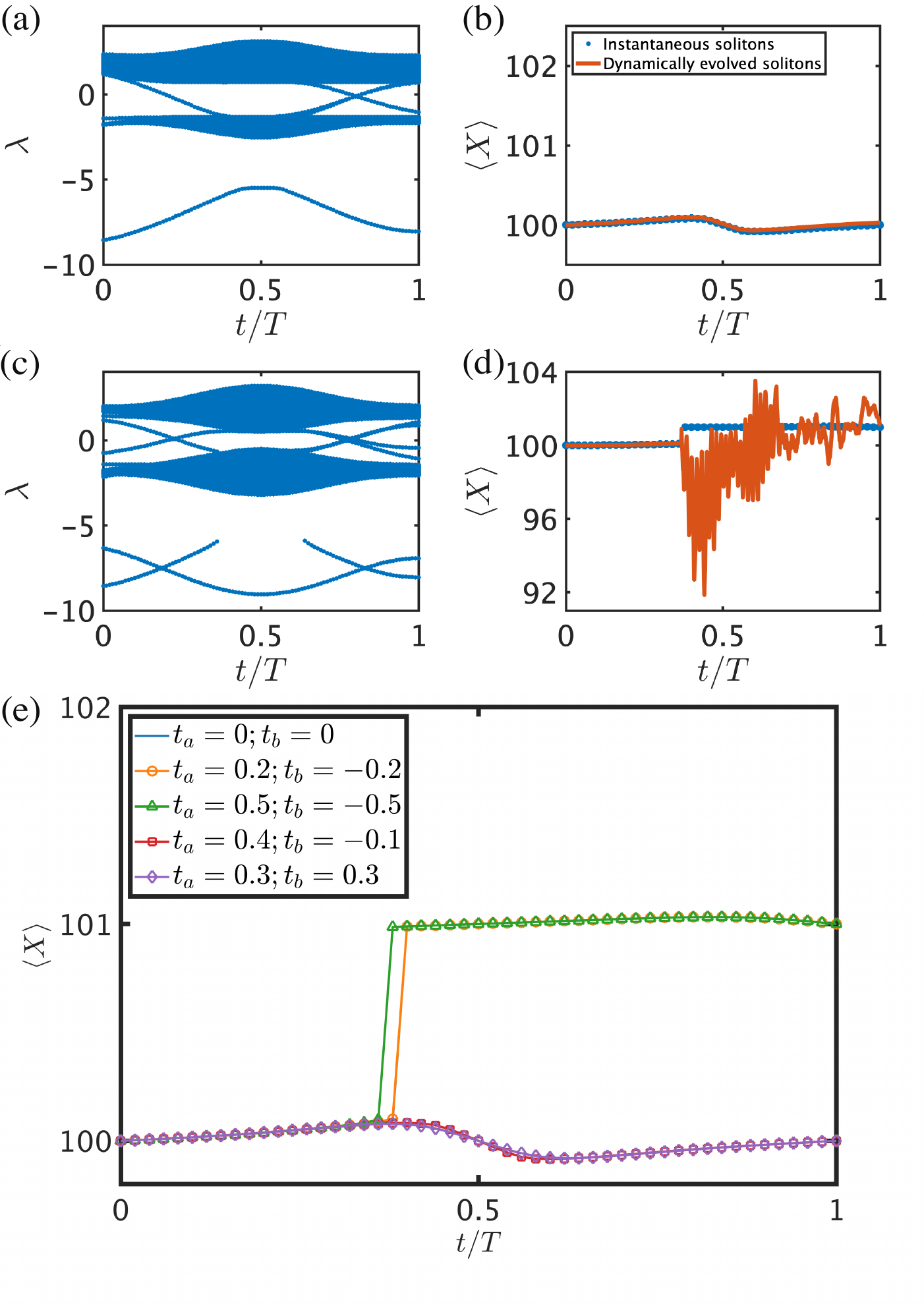}
\caption{\label{fig:7} Effects of parameter $t_a$ and $t_b$ on the auxiliary $\lambda$-spectrum defined in Eq. (\ref{eq:lambda}) for the nonlinear extended RM Hamiltonian (\ref{H}).The parameters are fixed
at $J=1$, $\delta=0.5$, $\Delta=1$, $g=7.5$, $T=2000\pi$, and $\omega = \omega_{\text{d}}=10^{-3}$. Other parameters are given as (a-b): $t_a=0.4$, $t_b=-0.1$; (c-d): $t_a=0.5$, $t_b=-0.5$. (e): Effects of parameter $t_a$ and $t_b$ on the anticipated position of the nonlinear excitation of soliton as a function of time over one period.}
\end{figure}

In the third scenario, we further investigate the influence of the NNN hopping terms $t_a$ and $t_b$ on the nonlinear eigenvalue system by choosing $g = 7.5$. It is noted that the nonlinearity strength $g$ at this point is higher than that set for the linear eigenvalues, due to the fact that the introduction of nonlinear eigenvalues causes the critical point of the nonlinearity strength $g$, influenced by $t_a$ and $t_b$ in the phase diagram, to increase.When $t_a = -t_b = 0.5$, the energy spectrum exhibits a self-intersecting structure of the ground-state soliton, as shown in Figs.~\ref{fig:7}(c). Correspondingly, in the phase diagram, adiabatic breakdown is observed, as depicted in Fig.~\ref{fig:7}(d). In contrast, when $t_a = 0.4$, $t_b = -0.1$, similar to the case with linear eigenvalues, the self-intersecting structure of the ground-state soliton disappears, and the pumping effect ceases, as shown in Figs.~\ref{fig:7}(a-b).

The comparative analysis in Figs.~\ref{fig:7}(e) reveals the parametric dependence of soliton dynamics: akin to the case with linear eigenvalues, adiabatic breakdown still occurs when $t_a = -t_b$; Whereas in other cases, the pumping effect is halted. It is noteworthy that, unlike the linear eigenvalue case, when $t_a = t_b = 0$, the phase diagram shows the cessation of the pumping effect, whereas in the linear eigenvalue scenario, adiabatic breakdown is observed. However, it should be noted that the case where $t_a = t_b = 0$ corresponds to the original nonlinear RM model, not the nonlinear extended RM model, and thus this phenomenon can be considered negligible. The above analysis indicates that the NNN terms do indeed selectively modify the system's phase diagram, especially when the parameter $t_a = -t_b$.

We remark that we use both dynamically evolved soliton method and instantaneous soliton solution method to calculate the expectation value $\langle X\rangle=\sum_j j|\Psi_j|^2$ as shown in Figs. \ref{fig:2}-\ref{fig:7}. In the first approach, we obtain the soliton at time $t$ by directly applying a given initial soliton and solving the equation for the time evolution of the wavefunction using the fourth-order Runge-Kutta method. In the second method, we determine the soliton at time $t$ through an iterative process, aiming to reach the steady state. Under adiabatic conditions, two methods converge as it is expected, providing mutual verification. Under non-adiabatic conditions related to self-intersecting bands (see Figs. \ref{fig:2}(c1), \ref{fig:4}(c), \ref{fig:5}(c1) and \ref{fig:7}(c1)) induced by the interaction, which result in discrepancies between instantaneous and dynamical soliton simulations elaborately (see Figs. \ref{fig:2}(c2), \ref{fig:4}(d), \ref{fig:5}(c2) and \ref{fig:7}(c2)), the first approach reflects physical reality in experimental implementations, while the second approach serves as a diagnostic tool for quantifying adiabaticity breakdown.

This section presents the energy spectra and phase diagrams for the nonlinear extended RM model under linear and nonlinear eigenvalue conditions. Comparative analysis shows that the nonlinearity strength $ g $ controls phase diagram evolution: increasing $ g $ drives transitions from perfect adiabatic transport through breakdown to complete suppression of pumping. The NNN parameters $ t_a $ and $ t_b $ modulate these phase transitions. Crucially, when $ t_a = -t_b $, the phase diagram exhibits characteristic restructuring that manifests as ground-state soliton reconfiguration in the energy spectrum.  We further observe fractional topological phases emerging at specific combinations of $ g $, $ t_a $, and $ t_b $. These phases display transport properties sensitive to eigenvalue variations, reflecting changes in topological invariants.

\section{Nonlinear bulk-edge correspondence}
\label{Section4}

In the preceding  Sec.~\ref{Section3}, we studied the nonlinear extended RM model under linear and nonlinear eigenvalue conditions, obtaining its $\lambda$ spectrum and ground-state pumping diagram. Furthermore, we determined the impact of NNN parameters $t_a$ and $t_b$ on the system's phase and energy spectra. Section~\ref{Section4} extends this investigation by analyzing how $t_a$ and $t_b$ influence anomalous nonlinear eigenvalue solutions in the extended RM Hamiltonian~(\ref{H}).

\begin{figure}
\includegraphics[width=1\columnwidth]{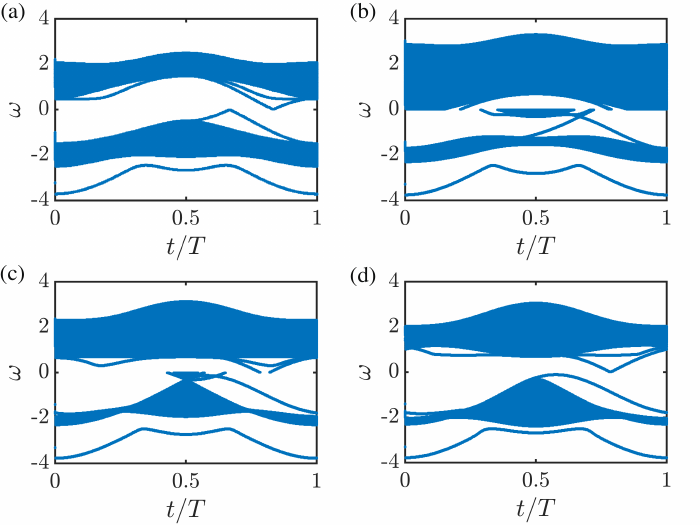}
\caption{\label{fig:8} Band structure $\omega$-$t$ of the auxiliary eigenvalue by solving Eq.~(\ref{eq:lambda}) at $\lambda = 0$. The parameters are fixed at $g=2$, $J=1$, $\delta=0.5$, $\Delta=1$, $T=2000\pi$, and $\omega_{\text{d}}=10^{-3}$. Parameters $t_a$ and $t_b$ are given as (a): $t_a=0$, $t_b=0$; (b): $t_a=0.4$, $t_b=0.4$; (c): $t_a=0.4$, $t_b=-0.1$; (d): $t_a=0.4$, $t_b=-0.4$.}
\end{figure}

First, we analyze how the NNN hopping parameters $t_a$ and $t_b$ affect nonlinearity in the anomalous eigenvalue problem for the extended Rice-Mele Hamiltonian (Eq.~\ref{H}). We specifically solve Eq.~(\ref{eq:lambda}) at $\lambda = 0$ to examine the relationship $t$--$\omega$. This is achieved by computing $t$--$\omega$ diagrams for varying $t_a$ and $t_b$ at fixed $\lambda=0$. Our analysis shows two key features: (i) Soliton states emerge irrespective of the specific values of $t_a$ and $t_b$, and (ii) increasing the interaction strength $g$ lowers the ground-state soliton energy. To clearly resolve two distinct soliton states, we focus our analysis on $g=2$.

Next, we plot the $\omega$-$t$ diagrams for different $t_a$ and $t_b$ values at $\lambda = 0$ and $g=2$ (Fig.~\ref{fig:8}). In the original model ($t_a = t_b = 0$; Fig.~\ref{fig:8}(a)), edge states display a sharp turning point. Setting $t_a = t_b = 0.4$ (Fig.~\ref{fig:8}(b)) broadens the bulk bands and introduces a new bulk band between them. For $t_a = 0.4$, $t_b = -0.1$ (Fig.~\ref{fig:8}(c)), bulk band widths decrease and the intermediate band diminishes. Here, the central soliton state connects to this intermediate band, becoming more distinct as bands narrow. The upper edge state retains its sharp turn, exhibiting fracture near $\omega=0$ in the latter half-period, while the lower state curves. When $t_a = -t_b = 0.4$ (Fig.~\ref{fig:8}(d)), band edges narrow at the upper bulk band while the lower band widens centrally, with complete disappearance of the intermediate band. This fully reveals the soliton band. Both edge states maintain their characteristic forms (sharp-angled upper, curved lower), with the fractured upper edge subsequently reconnecting.

At last, the NNN parameters $t_a$ and $t_b$ modify the anomalous eigenvalue solutions of the extended nonlinear RM model. The most significant modifications occur in the $\omega$-$t$ diagram when $t_a = -t_b$. Specifically:  (i) Nonzero $t_a$, $t_b$ values generate a new bulk band between existing bands;  (ii) When $t_a = -t_b$, this emergent band vanishes entirely.

\section{Conclusion and outlook}
\label{Section5}

This study focuses on investigating the behavior of the nonlinear extended RM model under both linear and nonlinear eigenvalue conditions. We systematically examine the energy spectra and phase diagrams of the system under these two distinct eigenvalue scenarios, with a particular emphasis on how the parameters $t_a$ and $t_b$ of the NNN terms influence the soliton dynamics and pumping transport. A key contribution of this work lies in transforming the problem of eigenfunction nonlinearity in Eqs. (\ref{GP1}) and (\ref{GP2}) into a problem of eigenvalue nonlinearity in Eq. (\ref{LinerP}). Notably, we find that assigning specific values to $t_a$ and $t_b$ gives rise to the emergence of fractional topological phases in the system—an observable physical phenomenon absent in conventional frameworks ($\hat{H}\Psi=E\Psi$) and unique to the nonlinear eigenvalue problem described by Eq. (\ref{LinerP}). Furthermore, when t
$ t_a $ and $ t_b $ , the range of nonlinearity strength enabling adiabatic breakdown broadens: the threshold for the onset of adiabatic breakdown decreases, whereas the threshold for its cessation increases. Additionally, we uncover a notable phenomenon: the nonlinear parameter $g$ modulates the system's topological invariants.

The study also establishes the bulk-edge correspondence in the anomalous nonlinear extended RM model. We find that the parameters $t_a$ and $t_b$ of the NNN terms also modify the system's $\omega - t$ diagram. Specifically, when $t_a = -t_b$, this diagram exhibits a striking modification: the bulk bands emerging from the NNN terms vanish entirely. This work provides key insights into the extended RM model's behavior under nonlinearity, revealing new phenomena at the interplay between nonlinear interactions and NNN couplings. These results have important implications for designing topological insulators and controlling edge states in nonlinear quantum systems, motivating further studies on nonlinear topological phenomena.

\textit{Acknowledgements} -- We thank Ying Hu, Yapeng Zhang, Xuzhen Cao, Shujie Cheng, and Biao Wu for stimulating discussions and useful help. We sincerely acknowledge Yucan Yan for his dedicated efforts during the critical inception phase of this project. This work was supported by the Zhejiang Provincial Natural Science Foundation (Grant No. LZ25A040004) and the National Natural Science Foundation of China (Grants No. 12574301).

\appendix
		
\section{Two methods for pumping path analysis\label{appendix A}}

In this Appendix \ref{appendix A}, we will provide a comprehensive description of the technical details pertaining to the computation of ground-state soliton pumping expectation values, encompassing both the instantaneous soliton and dynamical evolution methods~\cite{Bai2025,Tuloup2023}. Specifically, this will include:
(1) Numerical algorithms and parameter configurations for instantaneous soliton and dynamical evolution methods;
(2) Annotated citations of key references supporting the methodological framework.

In the first approach, designated as the method of dynamically evolved solitons, we implement a fourth-order Runge-Kutta algorithm to numerically propagate the initial soliton profile through temporal evolution governed by the nonlinear Schrödinger equations specified in Eqs. (\ref{GP1}) and (\ref{GP2}). This methodology faithfully reproduces the dynamical evolution of the wavefunction under physically realistic time-dependent driving conditions, thereby providing a direct computational realization of the quantum transport process. The numerical scheme preserves the symplectic structure of the underlying physical system while maintaining a balance between computational accuracy and efficiency, with the truncation error controlled at $O(\Delta t^4)$ through the fourth-order integration framework.

In the second approach called by the method of instantaneous solitons,  an iterative self-consistent scheme is employed to converge to the steady-state solution at each time slice $t$, bypassing explicit time propagation~\cite{Tuloup2023}. For a given nonlinear $\omega$-dependent Hamiltonian $H(\omega_{\text{d}},t)$, at a specific time t, the iterative procedure from state $|\Psi_{n}(t)\rangle$ to $|\Psi_{n+1}(t)\rangle$ proceeds as follows: 
\begin{itemize}
\item First, construct the nonlinear Hamiltonian $H_{n}$ based on the given state $|\Psi_{n}(t)\rangle$; 
\item Then, solve for the eigenstates $|\psi_{i}\rangle$of $H_{n}$ with $i=1, 2\dots 2N$; 
\item Finally, select the eigenstate with the maximum overlap with the previous state as the new state: $|\Psi_{n+1}(t)\rangle=|\psi_{i0}\rangle$ where $\langle \Psi_{n}(t)| \psi_{i0}\rangle \geq \langle \Psi_{n}(t)|\psi_{i}\rangle$ for all i. 
\end{itemize}
The iteration stops when the difference between the eigenvalues of the new state and the previous state is less than $\epsilon$ (in this work, $\epsilon=10^{-12}$ is chosen). It is crucial to note that the choice of the initial state significantly affects the final steady-state solution, hence the selection of the initial state requires extreme caution.

In the adiabatic regime ($\omega_{\text{d}} \rightarrow 0$), the two approaches converge to the same soliton trajectory, confirming numerical consistency. For nonadiabatic driving ($\omega_{\text{d}} \gg 0$), the first method faithfully reproduces experimentally observable dynamics, while the second method serves as a diagnostic tool to quantify deviations from adiabaticity by comparing instantaneous eigenstates with dynamically evolved states. We remark that, in the second approach, the wave function evolves through iterative steps as an iterative term, thereby ensuring the successful derivation of the band structure for the nonlinear wave function.

\bibliography{Reference}

\end{document}